\documentclass{aa}  

    \usepackage{graphicx}
    \usepackage{txfonts}
    \usepackage[colorlinks = true, allcolors=blue]{hyperref}
    \usepackage{stfloats}
    \usepackage{booktabs}
    \usepackage{orcidlink}
    \usepackage{sidecap}
    \usepackage{pdflscape}
    \newcommand\farcsec{\mbox{$.\!\!^{\prime\prime}$}}%
\begin{document}

       \title{The morphology and kinematics of the galaxy group AM\,1054-325: A MUSE perspective}
       % \subtitle{}
       \titlerunning{MUSE view of the galaxy group AM\,1054-325}
       % \authorrunning{Jadhav}
       
       \author{Jyoti Yadav \orcidlink{0000-0002-5641-8102}\inst{1,2},
              Vikrant V. Jadhav \orcidlink{0000-0002-8672-3300}\inst{3}}
    
       \institute{
            Indian Institute of Astrophysics, Koramangala II Block, Bangalore 560034, India\\
            \email{jyoti [at] iiap.res.in}
            \and
            Pondicherry University, R.V. Nagar, Kalapet, 605014, Puducherry, India
            \and
            Helmholtz-Institut für Strahlen- und Kernphysik, Universität Bonn, Nussallee 14-16, D-53115 Bonn, Germany
            }
    
       \date{Received May 18, 2024; accepted June 6, 2024}
    
    % \abstract{}{}{}{}{} 
    % 5 {} token are mandatory
     
      \abstract
      % context heading (optional)
       {
Galaxy interactions in groups can lead to intense starbursts and the activation of active galactic nuclei (AGNs). The stripped gas from the outer disk can lead to star-forming clumps along the tidal tails or sometimes tidal dwarf galaxies.
We investigate the impact of interaction on various galaxy properties, including morphology, star formation rates, and chemical composition in the galaxy group AM\,1054-325 using Multi Unit Spectroscopic Explorer (MUSE) data.
We condudt a comprehensive spatially and spectrally resolved investigation of the star formation rate, star formation histories, metallicity, and AGN activity.
The galaxy subgroup AM\,1054-325A shows multiple star-forming clumps in H$\alpha$ emission along the western tidal tail, which are formed due to tidal stripping. These clumps also have higher metallicities. AM\,1054-325B is quenched and shows disturbed gas kinematics and the signature of gas accretion in the H$\alpha$ map. 
The specific star formation along the tidal tail is higher, contributing to the galaxy's overall stellar mass growth.
       }
    
    
       \maketitle
    %
    %-------------------------------------------------------------------

\section{Introduction} \label{introduction}
The interaction of galaxies plays a crucial role in their evolution. Such interactions lead to the growth of supermassive black holes (SMBHs), bulges, and the formation of massive galaxies \citep{Matteo2005Natur.433..604D}. Galaxy groups, particularly those with abundant cold gas reservoirs, create highly favourable conditions for such activities due to the close interactions between galaxies. These interacting galaxies are classified based on their masses: major mergers with a mass ratio between the two galaxies greater than 1:4 and minor mergers with a mass ratio lower than 1:4. Collisions between major mergers are anticipated to be the most violent, while interactions involving minor companions are expected to be more common. This prevalence is attributed to the higher fractional abundance of low luminosity galaxies \citep{Lambas2012A&A...539A..45L}. The gravitational forces during galaxy interactions can lead to the collapse of interstellar gas and dust, triggering intense bursts of star formation and resulting in young, hot and massive stars. This has been extensively studied through observational data, which has revealed a clear connection between galaxy interactions and the initiation of star formation \citep{Yee1995ApJ...445...37Y, Kennicutt1998ARA&A..36..189K, Kauffmann2000MNRAS.311..576K, Tonnesen2012MNRAS.425.2313T}. Simulations also demonstrate that galaxy interactions lead to an elevated star formation rate (SFR) across the entire galaxy, including within the tidal tails formed due to the accretion, redistribution, and compression of gas induced by tidal interactions \citep{Renaud2015MNRAS.446.2038R, Moreno2021MNRAS.503.3113M}.

Tidal forces during galaxy interactions lead to the inflow of large amounts of gas and dust towards the central regions of the galaxies. This funnelled gas feeds the black hole and ignites active galactic nucleus (AGN) activity \citep{Hernquist1995ApJ...448...41H}. The majority of ultraluminous infrared galaxies and quasars (> 80\%) exhibit indications of a current or recent merger of galaxies \citep{Sanders1988ApJ...325...74S, Sanders1988ApJ...328L..35S, Urrutia2008ApJ...674...80U}. \citet{Ellison2013MNRAS.430.3128E} conducted a study on low-redshift major galaxy pairs (stellar mass ratio $<$ 4) selected from the Sloan Digital Sky Survey. They observed a clear pattern of increasing AGN excess (the ratio of the AGN fraction in paired galaxies compared to a control sample of isolated galaxies) as the projected separation between galaxies decreased (< 40 kpc). Similar approaches used in various studies of the nearby galaxies also show AGN enhancement in merging or interacting galaxies \citep{Alonso2007MNRAS.375.1017A, Ellison2011MNRAS.418.2043E, Satyapal2014MNRAS.441.1297S}.
When galaxies merge, the proximity between their SMBHs gradually decreases due to the effects of dynamical friction. If the SMBHs are in an active accretion phase, they can form dual AGNs \citep{Koss2012ApJ...746L..22K, Imanishi2020ApJ...891..140I} if they come closer. In galaxy groups, multiple galaxies can participate in the encounter process, potentially resulting in the formation of a triple AGN system \citep{Koss2012ApJ...746L..22K, Yadav2021A&A...651L...9Y}.

\begin{figure*}
    \centering
    \includegraphics[width=0.95\textwidth]{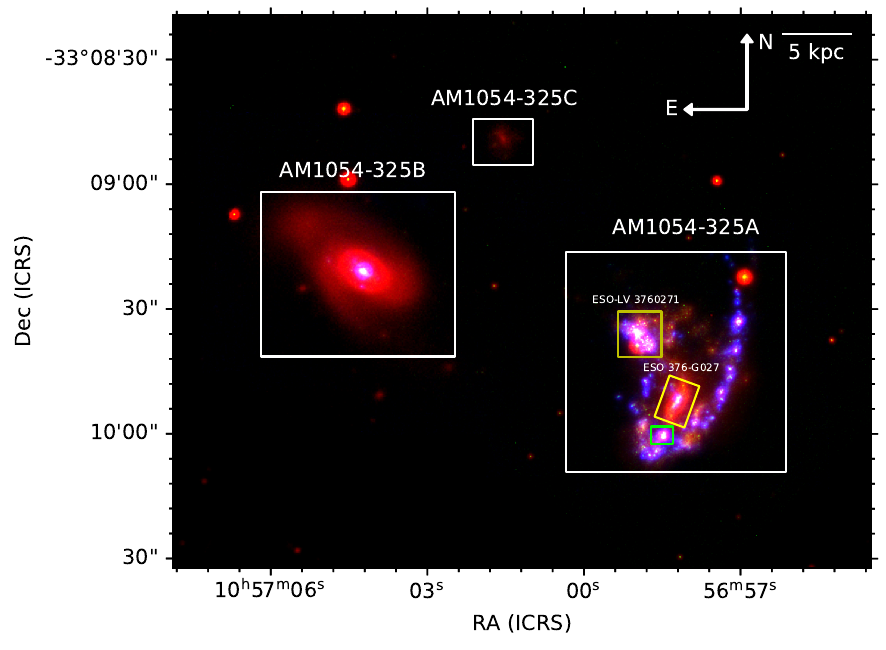} 
    \caption{Colour-combined image of the AM\,1054-325 galaxy group. The red colour represent z-band data from The Dark Energy Camera Legacy Survey (DECaLS), the green represents \textit{HST} F336W data, and the blue represents the H$\alpha$ emission. The tidal tail shows clumps that are bright in H$\alpha$ emission. White rectangles represent the galaxy AM1054-325B and the galaxy subgroup AM1054-325A. The yellow rectangles represent the merging galaxies ESO376-G027 and ESO-LV3760271, and another possible merging candidate galaxy AM\,1054-325D is shown in the green rectangle.}
    \label{fig:AM1054_rgb}
\end{figure*}

The gravitational forces and torques during the encounter of gas-rich galaxies can lead to the expulsion of neutral hydrogen gas, stars, and dust from the disk of these galaxies. This can result in various tidal features such as rings, tidal tails, bridges, and plumes \citep{Duc1999IAUS..186...61D}. These tidal tails and bridges provide valuable insights into the dynamic processes during galaxy interactions. In some cases, the material in the tidal tail can gravitationally collapse, forming tidal dwarf galaxies (TDGs; \citealt{Duc1999IAUS..186...61D}). These TDGs become kinematically detached from their host galaxies and exhibit characteristics similar to those of independent dwarf galaxy populations. Sometimes, two visually overlapping galaxies can be mistaken for interacting systems, making spectroscopic observations crucial for understanding the true nature of these objects \citep{Yadav2022A&A...657L..10Y}.

The interactions can also lead to gas accretion \citep{Balsara1994ApJ...437...83B, Acreman2003MNRAS.341.1333A, Rasmussen2006MNRAS.370..453R, Yadav2023MNRAS.526..198Y}. The accreted gas can settle in the outer disk of galaxies, as seen in extended UV disk galaxies, and can fuel star formation over extended periods \citep{Yadav2021ApJ...914...54Y}. The accreted gas can either be aligned or misaligned with the stellar rotation. When the accreted gas carries a significant amount of angular momentum of its own or has a different orbital trajectory, it can lead to misaligned/counter-rotating gas with respect to the stellar disk \citep{Corsini2014ASPC..486...51C}. The first observation of counter-rotation was reported by \citet{Galletta1987ApJ...318..531G}. They observed that in the SB0 galaxy NGC4546, the gas rotates in retrograde orbits compared to the stars. This peculiar kinematic misalignment has been observed not only between gaseous and stellar components \citep{Rubin1992ApJ...394L...9R, Bertola1996ApJ...458L..67B, Johnston2013MNRAS.428.1296J, Krajnovic2015MNRAS.452....2K}, but also between stellar components \citep{Bettoni1984Msngr..37...17B, Ciri1995Natur.375..661C, Pizzella2018A&A...616A..22P}.

The order of this paper is as follows. In Sect.~\ref{AM1054325}, we describe the galaxy group AM\,1054-325. Sect.~\ref{Muse} describes the Multi Unit Spectroscopic Explorer (MUSE) data and Galaxy IFU Spectroscopy Tool (\textsc{gist}) pipeline. In Sect.~\ref{analysis} and Sect.~\ref{conclusions}, we present the data analysis and summary.

\begin{table}
    \centering
    \small 
    \begin{tabular}{lcccr}
    % \caption{Details of the galaxy group AM\,1054-325}
    \toprule
    Source & $\alpha_{J2000}$ & $\delta_{J2000}$ & z  \\
    & (hh:mm:ss)& (dd:mm:ss) &    \\
    \hline
    AM\,1054-325A &          &             &        \\
    \quad\quad ESO\,376-G027 & 10:56:58 & $-$33:09:52 & 0.01264 \\ 
    \quad\quad ESO-LV 3760271 & 10:56:59 & $-$33:09:38 & 0.01287 \\
    AM\,1054-325B$^{\dagger}$ & 10:57:04 & $-$33:09:20 & 0.01284\\
    AM\,1054-325C$^{\ddagger}$ & 10:57:01 & $-$33:08:47  & --\\
    \toprule
    \end{tabular}
    \caption{Details of the galaxy group AM\,1054-325.}
    Notes: Columns represent the Names, location, and redshift of the galaxies within the galaxy group AM\,1054-325 respectively. ESO376-G027 and ESO-LV 3760271 belong to the galaxy subgroup AM\,1054-325A. $^{\dagger}$Also known as ESO376-G028. $^{\ddagger}$Out of the MUSE field of view. 
    \label{tab:details}
\end{table} 

%  \begin{figure*}
% \sidecaption
%   \includegraphics[width=12cm]{AM1054_segments_regions.pdf}
%      \caption{Background image representing the H$\alpha$  emission, with different colours indicating different segments from \textsc{SExtractor}. The cyan regions show the detected clumps overlaid on the H$\alpha$ map.}
%      \label{fig:Ha_sfcs}
% \end{figure*}

\begin{figure*}
    \centering
    \includegraphics[width=0.95\textwidth]{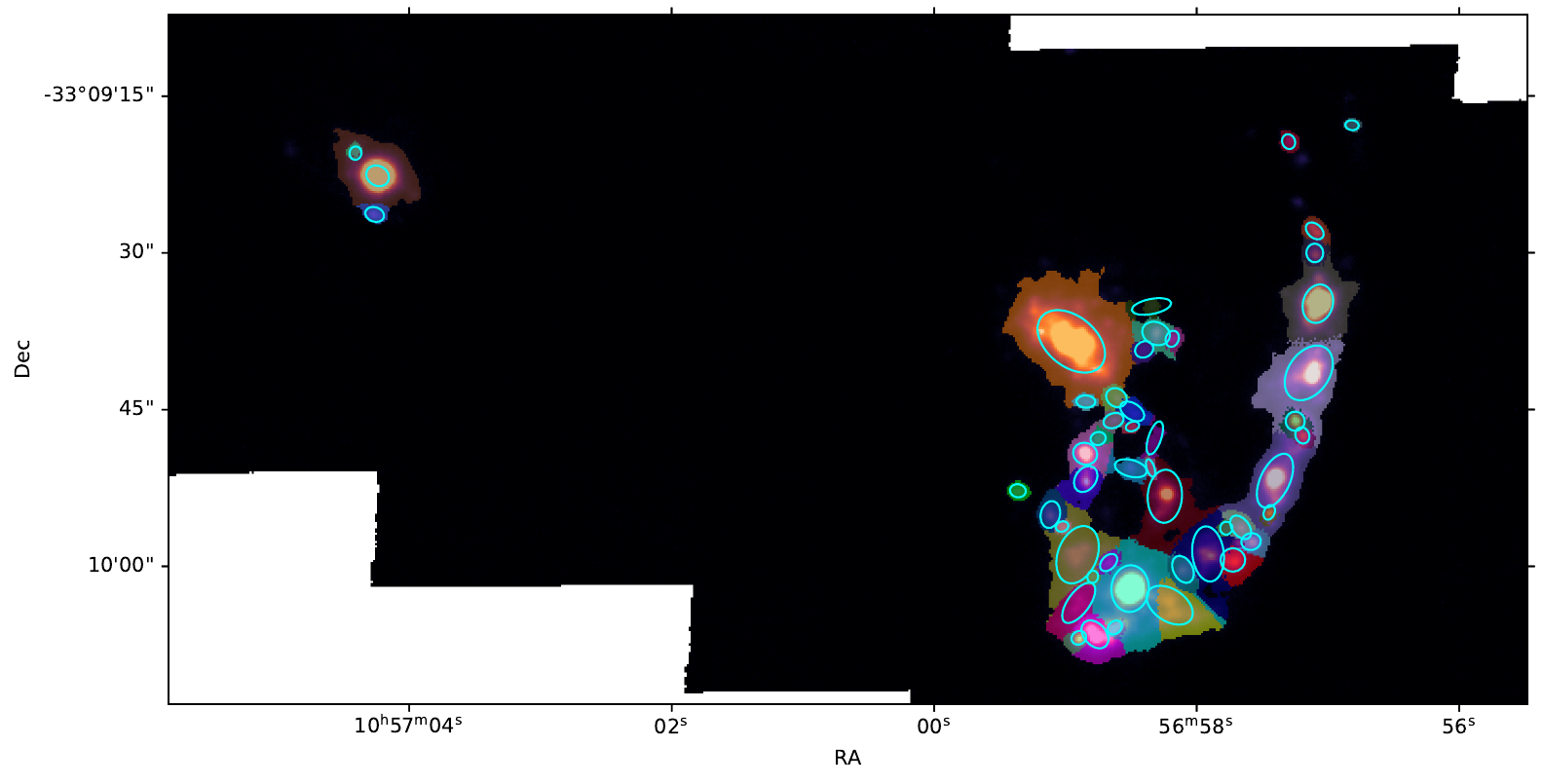} 
    \caption{Background image representing the H$\alpha$  emission, with different colours indicating different segments from \textsc{SExtractor}. The cyan regions show the detected clumps overlaid on the H$\alpha$ map.}
    \label{fig:Ha_sfcs}
\end{figure*}

\section{AM\,1054-325 Group} \label{AM1054325}
AM\,1054-325 is a galaxy group of multiple interacting galaxies. Fig.~\ref{fig:AM1054_rgb} shows the colour-combined image of the AM 1054-325 galaxy group. The AM\,1054-325A galaxy subgroup is composed of multiple arms and shows a peculiar morphology, while the other galaxy, AM\,1054-325B, is a spiral galaxy. The spiral galaxy AM\,1054-325B, also known as ESO376-G028, is optically bright. AM\,1054-325A is composed of two nuclei,  ESO-LV 376-0271 and ESO376-G027, which are merging. The AM\,1054-325A galaxy subgroup also shows a long tidal tail containing star-forming clumps, which exhibit beads-on-string morphology. These clumps may be tidal dwarf galaxies in the process of formation. \citet{Rosa2014MNRAS.444.2005R} showed that the emission lines from these HII regions are excited by shocks. AM\,1054-325B also appears to interact with another low-surface-brightness galaxy, AM\,1054-325C. However, the redshift of AM\,1054-325C is unknown. The details of the galaxy group are given in Table \ref{tab:details}.

\section{Data} \label{Muse}
MUSE is an integral field unit (IFU) on the Yepun Very Large Telescope (VLT). MUSE provides 3D imaging and spectroscopic data over a wavelength range of 4650--9300 \AA. 
It operates in two modes: the wide-field mode (WFM) and the narrow-field mode. We used data from the WFM, which offers a field of view of 1$\arcmin\times$1$\arcmin$ with 0\farcsec2 sampling. MUSE uses the GALACSI adaptive optics module to provide adaptive optics corrected data, achieving a MUSE spatial resolution of 0\farcsec4 at 7000 \AA. MUSE has a resolving power of 1770 at 4800 \AA\ and 3590 at 9300 \AA. AM\,1054-325 was observed under Programme ID 106.2155 for a total exposure time of 7920 s. The total sky coverage was 4 arcmin$^{2}$ with a sky resolution of 1.2\farcsec

We used the \textsc{gist} version 3.0.3\footnote{\url{https://abittner.gitlab.io/thegistpipeline/}} \citep{2019Bittner} pipeline to analyse the MUSE data. The \textsc{gist} pipeline is designed to analyse fully reduced IFU data. For the analysis of emission lines, \textsc{gist} uses gas and absorption-line fitting (\textsc{GandALF}) \citep{2006Sarzi, 2006Falcon, 2019Bittner}. The penalised pixel-fitting (\textsc{pPXF}; \citealt{2004Cappellari,2017Cappellari}) method is used for stellar continuum fitting in \textsc{gist}. Data below a signal-to-noise ratio (S/N) of three were masked, and Voronoi binning was performed based on H$\alpha$ emission for an S/N of 50 for the gas emission analysis.
The galaxy group AM\,1054-325 shows massive and ongoing star-forming regions that are bright in H$\alpha$, resulting in smaller and finer bins. For stellar velocity and star formation history (SFH) estimation, we performed Voronoi binning for a wavelength range of 4765-5530 \AA. We corrected the spectra for the Milky Way extinction and fitted the stellar continuum using a multiplicative eighth-order Legendre polynomial.

\section{Analysis} \label{analysis}
\subsection{Star-forming regions} \label{sec:sfr_estimation}
A galaxy's young and massive star formation emits a significant amount of H$\alpha$ emission. H$\alpha$ emission traces the star formation up to 10 Myr. In Fig.\ref{fig:AM1054_rgb}, the blue shows the H$\alpha$ emission of the galaxy group. The AM\,1054-325A subgroup shows significant star formation in the clumps along the tidal tail. This suggests that interactions have led to tidal stripping of gas, followed by star formation. The H$\alpha$ emission also suggests that galaxy interactions have recently occurred, producing massive, young stars. AM\,1054-325B appears relatively quiescent in terms of star formation, exhibiting a comparatively subdued level of H$\alpha$ emission. The tidal tails associated with AM\,1054-325A exhibit massive clumps that are bright in H$\alpha$ emission. This difference in activity highlights the diverse nature of galaxies within this group. We used the Python library for Source Extraction and Photometry (\textsc{SExtractor}; \citealt{Bertin1996}) to identify and extract the clumps in the galaxies. \textsc{SExtractor} can execute various functions, including source detection, background estimation, and deblending. \textsc{SExtractor} identifies the sources based on a given threshold. Pixels with counts more than the threshold are identified as sources. These detected objects are then deblended and cleaned. We used the detection threshold (thresh) of 5$\sigma$, where $\sigma$ is the global background noise (i.e. the RMS  of the counts). The minimum number of pixels required for the identification of a region was fixed at 40 (minarea = 40). Thus, we detected regions with an area equivalent to the resolution of the MUSE data. We used elliptical shapes for the star-forming regions. We deblended the regions with 64 sub-thresholds (deblend\_nthresh = 64) and a minimum contrast parameter value of 0.00001 (deblend\_cont = 0.00001). We detected 48 clumps, three of which are galaxy cores. We also extracted the segmentation map, which gives the member pixels for each object. Fig.~\ref{fig:Ha_sfcs} shows the extracted clumps in cyan and segments represented by different colours over-plotted on H$\alpha$ emission.

\subsection{Area and star formation rate}
\begin{figure}
    \centering
\includegraphics[width=0.48\textwidth]{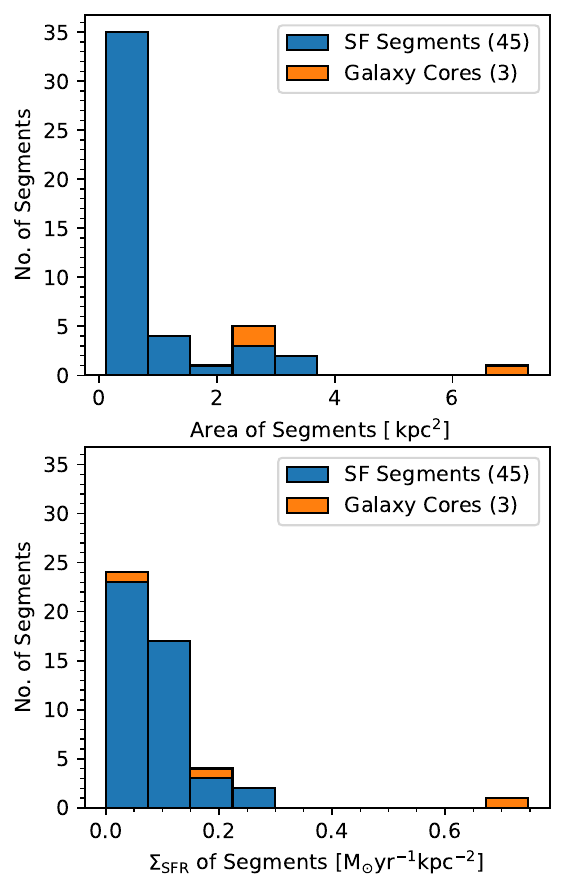}
    \caption{The figure represents the stacked histogram for the detected segments. The top and bottom panels show the stacked histograms of the area and $\Sigma_{SFR}$ for the detected segments (blue) and galaxy cores (orange) in H$\alpha$.}
    \label{fig:sfr_area}
\end{figure}

{We estimated the area corresponding to each segment and the corresponding SFRs.
%We also estimated each segment's SFR by adding the counts associated with each segment.
} 
We corrected for the Milky Way extinction using the Fitzpatrick law \citep{fitzpatrick1999}, and for internal extinction using the Balmer decrement. We assumed a temperature of 10$^{4}$ K and an electron density of 10$^{2}$ cm$^{-3}$ for Case B recombination, which corresponds to $(H_{\alpha}/H_{\beta})_{int}$ = 2.86 \citep{Osterbrock1989SvA....33..694O, Osterbrock2006agna.book.....O, Momcheva2013AJ....145...47M} This choice, which is standard for star-forming galaxies in the literature. The following equation gives the nebular colour excess: 
\begin{equation}
    E(B-V) = 1.97 \log_{10}\Biggr[\frac{(H_{\alpha}/H_{\beta})_{obs}}{2.86}\Biggr],
\end{equation}
where H$_{\alpha}$ and H$_{\beta}$ are the observed fluxes of H$_{\alpha}$ and H$_{\beta}$ emission lines, respectively.
We calculated the SFR surface density ($\Sigma_{SFR}$) for each segment in H$\alpha$ using the following relation \citep{Calzetti2007ApJ...666..870C}:
\begin{equation}
\label{eqn:sfrha}
SFR(H\alpha)  = 5.3\times10^{-42} L_{H\alpha},
\end{equation} 
where SFR(H$\alpha$) is the SFR in H$\alpha$ [M$_\odot$ yr$^{-1}$] and L$_{H\alpha}$ is the H$\alpha$ luminosity [erg s$^{-1}$].
{Fig.~\ref{fig:sfr_area} shows the histograms of the area and SFR/area ($\Sigma_{SFR}$) of the segments and galaxy cores.} 

\begin{figure*}
    \centering
    \includegraphics[width=0.9\textwidth]{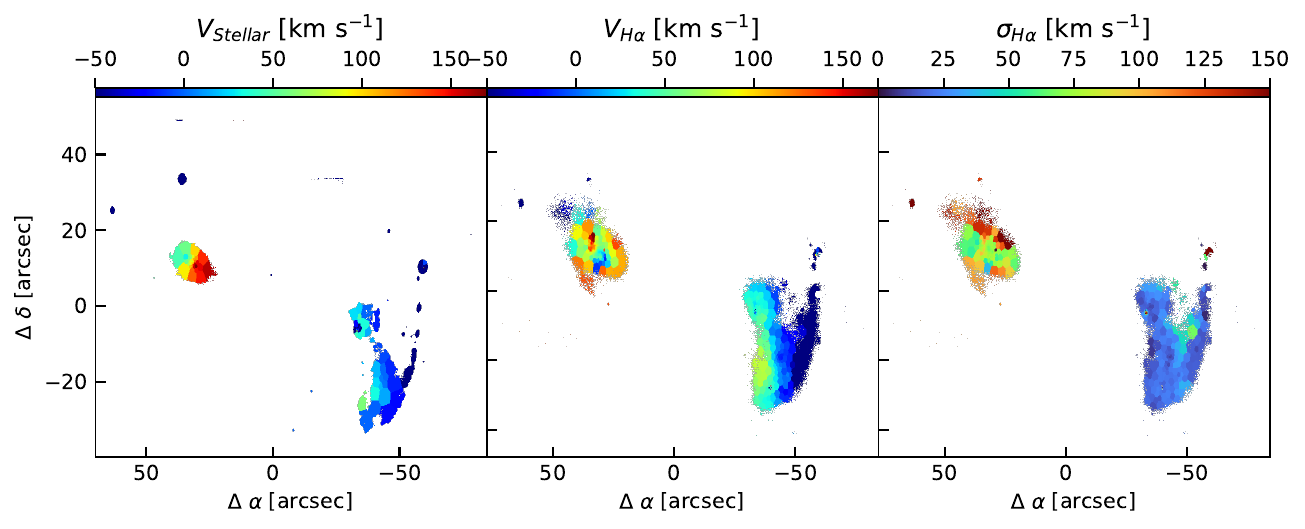} 
    \caption{The stellar and gas kinematics derived from \textsc{gist}. The left and right panels show respectively the stellar and H$\alpha$ gas velocities of the galaxy group. The galaxy AM\,1054-325B displays counter-rotating gas and stellar disk.}
    \label{fig:stellar_ha_vel}
\end{figure*}

\begin{figure}
      \centering
\includegraphics[width=0.48\textwidth]{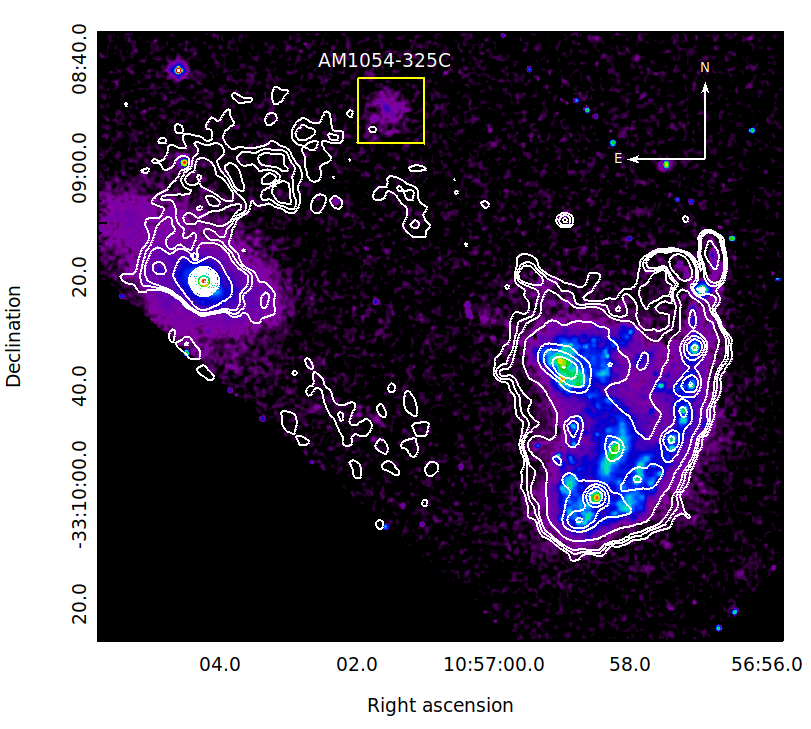} 
    \caption{\textit{HST} image with overlaid H$\alpha$ contours. The contours extend from AM1054-325B towards the north-west. This shows the signature of accreted ionised gas in AM\,1054-325B.}
    \label{fig:accretion}
\end{figure}

\begin{figure}
  \includegraphics[width=0.40\textwidth]{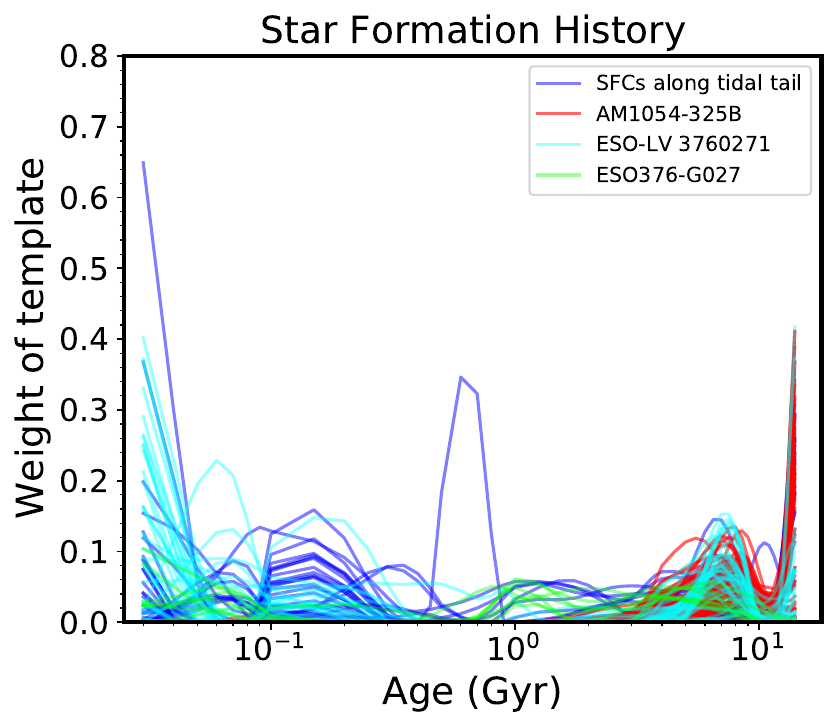}\\
   \includegraphics[width=0.40\textwidth]{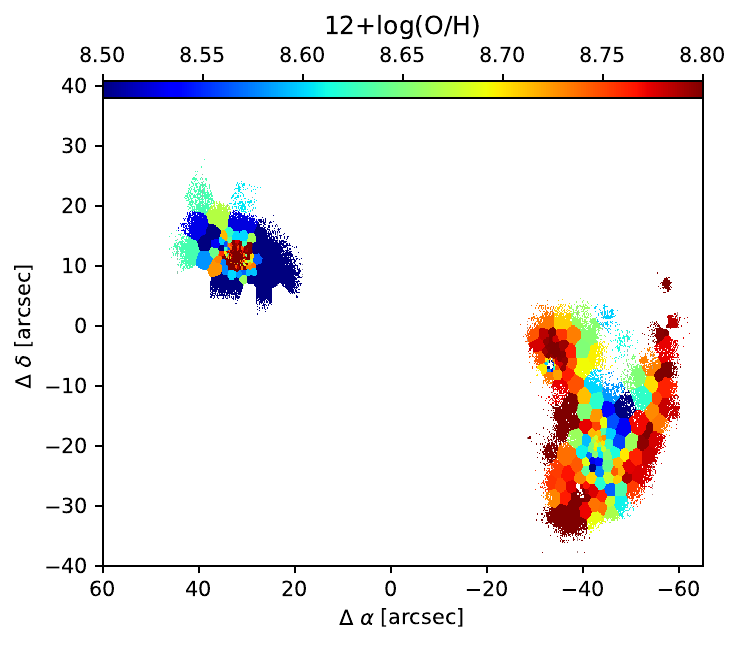}
     \caption{The top panel shows the star formation history, and the bottom panel shows the metallicity of the galaxy group. Bins with an amplitude-to-noise ratio (A/N) of more than 4 in all the H$\beta$, \ion{O}{iii}, H$\alpha$ \ion, and {N}{ii} lines were considered to be a reliable estimation of metallicity.}
     \label{fig:sfh_metallicity}
\end{figure}

% \begin{figure*}
%     \centering
%     \includegraphics[width=0.4\textwidth]{AM1054_sfh_2.pdf} 
%     \includegraphics[width=0.42\textwidth]{metallicity_2.pdf}  
%     \caption{The left panel shows the star formation history, and the right panel shows the metallicity of the galaxy group. Bins with an amplitude-to-noise ratio (A/N) of more than 4 in all the H$\beta$, \ion{O}{iii}, H$\alpha$ \ion, and {N}{ii} lines were considered to be a reliable estimation of metallicity.}    
%     \label{fig:sfh_metallicity}
% \end{figure*}

\begin{figure}
    \centering
    \includegraphics[width=0.48\textwidth]{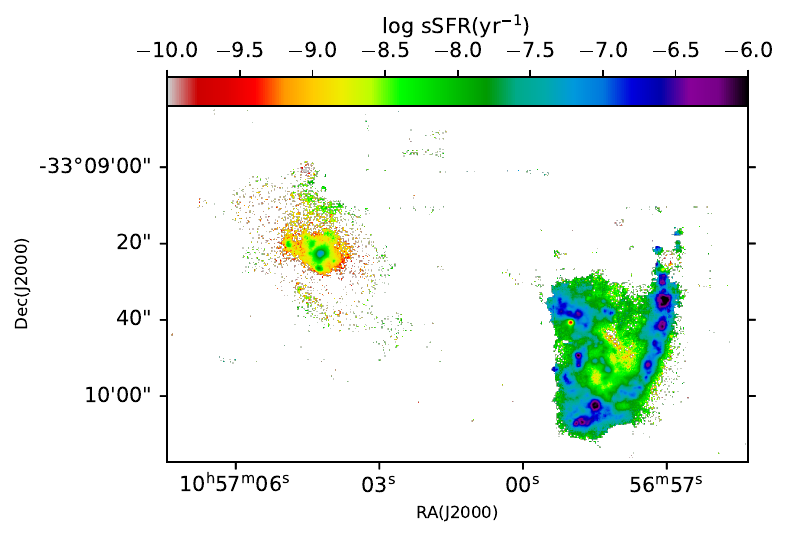} 
    \caption{sSFR distribution in the galaxy group AM\,1054-325.}
    \label{fig:ssfr}
\end{figure}

\subsection{Gas and stellar velocity}
The accreted external gas during galaxy interactions can have angular momentum that may not be perfectly aligned with the pre-existing stellar disk, which results in kinematic misalignments between the gas and stellar disks. In some cases, the accreted gas orbits in the opposite direction to the stellar disk, leading to counter-rotation between the gas and the stellar disks. We derived the H$\alpha$ and stellar velocity using the \textsc{gist} pipeline to understand the effect of interaction on the kinematic properties of the AM\,1054-325 galaxy group. The \textsc{gist} pipeline uses \textsc{GandALF} to estimate the gas emission line properties and \textsc{pPXF} to estimate the stellar kinematics. The left, middle, and right panels of Fig.~\ref{fig:stellar_ha_vel} show the stellar velocity, gas velocity, and gas velocity dispersion, respectively. The gas disk is kinematically disturbed relative to the stellar disk in AM\,1054-325B. Fig.~\ref{fig:accretion} shows the \textit{Hubble Space Telescope} (\textit{HST}) maps with overlaid H$\alpha$ intensity contours.
These contours extend from the north-west and south-west towards AM\,1054-325B. This indicates the presence of accreted ionised gas in AM\,1054-325B. The gas velocity map (Fig.~\ref{fig:stellar_ha_vel}, middle panel) also shows redshifted and blueshifted velocities with respect to the stellar disk in the northern and southern directions, suggesting an inflowing gas. The gas layers are more easily disturbed by tidal interactions than stellar disks. The gas velocity dispersion is also higher in the outskirts of AM\,1054-325B (Fig.~\ref{fig:stellar_ha_vel}, right panel). The star formation activity in AM\,1054-325B is significantly lower, indicating insufficient gas content to dissipate the momentum of the incoming accreted gas. Thus, the accreted gas settles into a disk rotating opposite to the stellar disk. Such accretion events result in gas replenishment, contributing to the extensive gas reservoirs surrounding most spiral galaxies. Under quiescent conditions, the gas slowly spirals inwards. However, initial tidal interactions can drive the gas towards the centre, significantly impacting the abundance gradients \citep{Montuori2010A&A...518A..56M}. In some cases, these gradients can even undergo reversal, as observed in the MASSIV survey \citep{Queyrel2012A&A...539A..93Q} at high redshifts. This reversal involves low-metallicity gas flowing into the centre and diluting the abundance of the central gas. Low-metallicity gas is observed in the outer regions of AM\,1054-325B (Fig.~\ref{fig:sfh_metallicity}, bottom panel), suggesting a possible connection to the accreted gas.

\subsection{Star formation history and metallicity}
The SFH module in the \textsc{gist} pipeline uses emission-line-subtracted spectra generated by the \textsc{GandALF} module to estimate the non-parametric SFH. The non-parametric SFH is calculated by modelling the observed spectrum using \textsc{pPXF}. The \textsc{gist} pipeline produces a linear combination of spectral templates and assigns them linear weights to match the model and observed spectra. These linear weights assigned to the template spectra are used to estimate the SFH and stellar population parameters. We estimated the SFH in AM\,1054-325 using \textsc{gist}. We also measured the gas phase metallicity of AM\,1054-325 using the flux of the emission lines provided by \textsc{gist}. The top panel of Fig.~\ref{fig:sfh_metallicity} shows the SFH map of AM\,1054-325. Metallicity maps were derived using the following equation from \citet{Pettini2004MNRAS.348L..59P}.
\begin{equation}\label{eq:metallicity}
    12+log\left(\frac{O}{H}\right)= 8.73 - 0.32\times
   \log\Biggr[ \frac{\ion{O}{iii}/H\beta}{\ion{N}{ii}/H\alpha} \Biggr],
\end{equation}

where \ion{O}{iii}, H$\beta$, \ion{N}{ii}, and H$\alpha$ are the \ion{O}{iii}, H$\beta$, \ion{N}{ii}, and H$\alpha$ emission line fluxes, respectively.

The top panel of Fig.~\ref{fig:sfh_metallicity} shows the SFH of each galaxy component in the galaxy group. The galaxy group shows multiple peaks in the SFH. ESO-LV 3760271 and the star-forming clumps (SFCs) along the tidal tail indicate recent star formation activity in the SFH plot, while AM\,1054-325B and ESO376-G027 exhibit star formation peaks in the past. The recent burst of star formation activity is also evident from the H$\alpha$ map, which shows bright clumps in the tidal tail. The metallicity of these clumps is also higher ({Fig.~\ref{fig:sfh_metallicity}, bottom panel). The metallicity map indicates that ESO376-G027 has a lower metal content than both its tail and ESO-LV 3760271. The tail that stretches out is connected to a core (shown in the green rectangle in Fig.~\ref{fig:AM1054_rgb}) beneath ESO376-G027. This core looks redder in the colour-combined image (Fig.~\ref{fig:AM1054_rgb}) and has a metallicity similar to the tidal tail (Fig.~\ref{fig:sfh_metallicity}). This suggests that the core might be the central, bulging part of another galaxy, AM\,1054-325D, that has been pulled away along the tail.

\begin{figure*}
    \centering
    \includegraphics[width=0.98\textwidth]{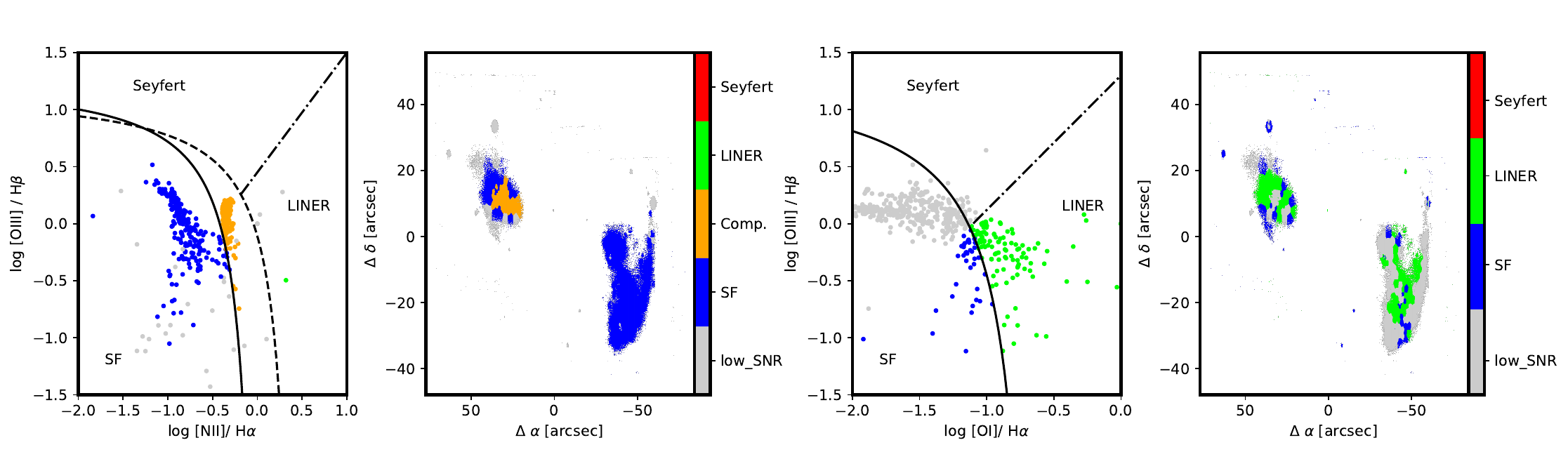} 
    \caption{BPT diagrams of the galaxy group AM\,1054-325. The left and right panels show the [\ion{N}{ii}] and  [\ion{O}{i}] BPT diagrams. Bins with an amplitude-to-noise ratio (A/N) of less than 4 in any of the lines are shown in grey.}
    \label{fig:BPT}
\end{figure*}

\subsection{Specific star formation rate}
The specific SFR (sSFR) quantitatively measures the contribution of star formation to galaxy growth and its relationship with stellar mass and the SFR. It is defined as the SFR per unit stellar mass. Several observational studies have established a correlation between SFR and stellar mass in star-forming galaxies.
To estimate the sSFR, we created an SFR map of the interacting group using the H$\alpha$ map. The stellar mass map was created using the method given in \citet{Bell2003ApJS..149..289B}: 
\begin{equation}
    log10(M/L_{i})= a_{i}+b_{i}\times(r-i),
\end{equation}
where the values of a and b are 0.006 and 1.114, respectively.
We used \textsc{pyMuse} \citep{Pessa2018arXiv180305005P} to create \textit{r}- and \textit{i}-band images of the galaxy group from MUSE cubes. We converted the images into flux maps (erg sec$^{-1}$ cm$^{-2}$ A$^{-1}$) by dividing the image by the central wavelength of the filter. We then converted the flux maps to magnitude maps and \textit{i}-band flux map to luminosity map. Pixels with values above 27 magnitudes were removed to eliminate noisy data.  Using the above formula, we then derived a pixel-wise stellar mass map. By dividing the SFR map by the stellar mass map, we generated a pixel-wise sSFR map. The sSFR of the galaxies is shown in Fig.~\ref{fig:ssfr}. AM\,1054-325A exhibits an elevated sSFR within both the tidal tail and ESO-LV 3760271 regions, whereas the sSFR of ESO376-G027 appears diminished, indicative of its quenched state. The sSFR of the tidal tail is different from that of ESO376-G027, and similar to that of the nucleus below ESO376-G027, which again supports the idea that the clumps in the tidal tail were not formed from ESO376-G027. These clumps were instead formed from the gas stripped from another galaxy, the nucleus of which is present below ESO376-G027.
The increase in the sSFR in the tidal tail implies a substantial contribution of ongoing star formation to the galaxy's overall stellar mass growth. This suggests the crucial role of star-forming processes in shaping the evolution of AM\,1054-325A.

\subsection{BPT diagram}
The investigation of AGN and star formation activity in interacting galaxies is essential as it provides insights into the influence of galaxy interactions on stimulating star formation and fuelling SMBHs. To elucidate the ionisation mechanism of the gas, the Baldwin-Phillips-Terlevich (BPT) diagram, as introduced by \citealt{1981Baldwin}, categorises sources based on the flux ratios of their emission lines. We generated BPT diagrams to investigate the characteristics of the emission mechanism. The BPT diagram utilises the [\ion{O}{iii}]$\lambda 5007/$H$\beta$ and [\ion{N}{ii}]$\lambda 6563/$H$\alpha$ line ratios as discerning indicators to distinguish and categorise different regions, such as Seyfert, low-ionisation nuclear emission-line region (LINER), composite, and star-forming regions. Fig.~\ref{fig:BPT} shows the BPT diagram representing the galaxy group AM\,1054-325. The emission detected from both the tidal tail and ESO-LV 3760271 regions falls within the star-forming region of the BPT diagram, while AM\,1054-325B exhibits composite emission. The stacked spectrum from some of the bins confirms the high level of [\ion{O}{i}] emission, which lies in the LINER part of the [\ion{O}{i}] BPT diagram. Since we only studied bins with A/N > 4, the high [\ion{O}{i}]/H$\alpha$ line ratios are unlikely to be due to measurement errors. The regions characterised by elevated [\ion{O}{i}]/H$\alpha$ ratios are primarily ionised by field OB stars, which emit photons from HII regions or by shocks. These shocks can originate from processes such as gas accretion and/or stripping or stellar feedback resulting from recent starburst activity. The redshifted and blueshifted velocities (Fig~\ref{fig:stellar_ha_vel}) show the gas flowing towards the inner regions. This inflowing gas collides with the interstellar medium of the host galaxy and can produce powerful shocks. The interaction processes in galaxy groups can produce tidal stripping and shock-heating of the gas, leading to LINER emission.  
A comprehensive study using  IFU data revealed a substantial proportion of shock excitation induced by tidal forces in nearby luminous infrared galaxies \citep{Monreal2010A&A...517A..28M}. In studies by \citet{Farage2010ApJ...724..267F} and \citet{Rich2010ApJ...721..505R, Rich2011ApJ...734...87R}, advanced models were used to analyse the shocked gas. In all cases, the shock excitation shows features similar to extended LINER-like emission with widened line profiles. These shocks result from substantial movements of gas triggered by the merger process. During a major merger, tidal forces and accretion processes drive gas inwards, causing shock excitation \citep{Farage2010ApJ...724..267F}. This infalling gas fuels significant increases in star formation and AGN activity, leading to massive galactic outflows and additional shocks in the interstellar medium and beyond.

%
%                                                One column figure
%----------------------------------------------------------------- 

%-----------------------------------------------------------------

\section{Summary} \label{conclusions}
This study uses archival data from the MUSE to present the properties of interacting galaxy group AM\,1054-325. MUSE's high spatial and spectral resolution played a crucial role in identifying the clumps and understanding their properties.
\begin{enumerate}
    \item The galaxy group AM\,1054-325 is composed of two subgroups, AM\,1054-325A and AM\,1054-325B. AM\,1054-325A shows massive star-forming clumps, which are formed due to tidal stripping.
    \item The clumps along the tidal tail are brighter in H$\alpha$, which suggests that they have formed recently: these clumps also have higher metallicities.
    \item AM\,1054-325B shows disturbed gas kinematics, possibly due to gas accretion onto the disk.
    \item The galaxy group shows multiple episodes of star formation; ESO-LV3760271 and SFCs along the tail show recent episodes of star formation activity.
    \item The sSFR is higher along the tidal tails, leading to the growth of stellar mass in the galaxy.
    \item The BPT diagram indicates ionisation due to shocks, which suggests that interaction and/or mergers can lead to shock heating of the gas.
\end{enumerate}

\begin{acknowledgements}
We thank the anonymous referee for the thoughtful review, which improved the impact and clarity of this work. VJ acknowledges support from the Alexander von Humboldt Foundation. This paper has used the observations collected at the European Southern Observatory under ESO programme 106.2155. This research has also used data from DECaLS at CTIO. This publication has used the NASA/IPAC Extragalactic Database (NED), operated by the Jet Propulsion Laboratory, California Institute of Technology, under contract with the National Aeronautics and Space Administration.
\end{acknowledgements}
\bibliographystyle{aa} % style aa.bst
\bibliography{sample}

\end{document}